\input harvmac\skip0=\baselineskip
\input epsf

\newcount\figno
\figno=0
\def\fig#1#2#3{
\par\begingroup\parindent=0pt\leftskip=1cm\rightskip=1cm\parindent=0pt
\baselineskip=11pt \global\advance\figno by 1 \midinsert
\epsfxsize=#3 \centerline{\epsfbox{#2}} \vskip 12pt {\bf Fig.\
\the\figno: } #1\par
\endinsert\endgroup\par
}
\def\figlabel#1{\xdef#1{\the\figno}}
\def\encadremath#1{\vbox{\hrule\hbox{\vrule\kern8pt\vbox{\kern8pt
\hbox{$\displaystyle #1$}\kern8pt} \kern8pt\vrule}\hrule}}



\lref\osv{
  H.~Ooguri, A.~Strominger and C.~Vafa,
  ``Black hole attractors and the topological string,''
  Phys.\ Rev.\ D {\bf 70}, 106007 (2004)
  [arXiv:hep-th/0405146].
}

\lref\WittenKT{
  E.~Witten,
  ``Three-Dimensional Gravity Revisited,''
  arXiv:0706.3359 [hep-th].
}

\lref\Tuite{
  M.~P.~Tuite,
  ``Genus Two Meromorphic Conformal Field Theory,''
  arXiv:math.qa/9910136.
}

\lref\Igusa{
  J.-I.~Igusa,
  ``On Siegel Modular Forms of Genus Two,"
  Am.J.Math. {\bf 84} (1962) 175-200;
  ``Modular Forms and Projective Invariants,"
  Am.J.Math. {\bf 89} (1967) 817-855.
}

\lref\FLM{ I.~Frenkel, J.~Lepowsky and A.~Meurman, ``A Natural
Representation of the Fischer-Griess Monster with the Modular
Function J as Character,'' Proc.Natl.Acad.Sci.USA {\bf 81} (1984)
3256-3260;
  I.~Frenkel, J.~Lepowsky and A.~Meurman,
  ``Vertex Operator Algebras and the Monster,''
{\it  Boston, USA: Academic (1988) 508 P. (Pure and Applied
Mathematics, 134)} }

\lref\ZamolodchikovAE{
  A.~B.~Zamolodchikov, ``Conformal Scalar Field on the
  Hyperelliptic Curve and Critical Ashkin-Teller Multipoint
  Correlation Functions,'' Nucl.\ Phys.\  B {\bf 285}, 481 (1987).
}

\lref\KnizhnikXP{
  V.~G.~Knizhnik,
  ``Analytic Fields on Riemann Surfaces. 2,''
  Commun.\ Math.\ Phys.\  {\bf 112}, 567 (1987).
}

\lref\DixonQV{
  L.~J.~Dixon, D.~Friedan, E.~J.~Martinec and S.~H.~Shenker,
  ``The Conformal Field Theory Of Orbifolds,''
  Nucl.\ Phys.\  B {\bf 282}, 13 (1987).
}

\lref\HamidiVH{
  S.~Hamidi and C.~Vafa,
  ``Interactions on Orbifolds,''
  Nucl.\ Phys.\  B {\bf 279}, 465 (1987).
}

\lref\MaldacenaBW{
  J.~M.~Maldacena and A.~Strominger,
  ``AdS(3) black holes and a stringy exclusion principle,''
  JHEP {\bf 9812}, 005 (1998)
  [arXiv:hep-th/9804085].
}

\lref\Hohn{ G.~Hoehn, ``Selbstduale Vertexoperatorsuperalgebren und
das Babymonster,'' arXiv:0706.0236. }

\lref\Tuiteb{ G.~Mason and M.~P.~Tuite, ``On Genus Two Riemann
Surfaces Formed from Sewn Tori,'' arXiv:math.qa/0603088. }

\lref\ManschotZB{
  J.~Manschot,
  ``$AdS_3$ Partition Functions Reconstructed,''
  arXiv:0707.1159 [hep-th].
}

\lref\GaiottoXH{
  D.~Gaiotto and X.~Yin,
  ``Genus Two Partition Functions of Extremal Conformal Field Theories,''
  arXiv:0707.3437 [hep-th].
}

\lref\KrasnovZQ{
  K.~Krasnov,
  ``Holography and Riemann surfaces,''
  Adv.\ Theor.\ Math.\ Phys.\  {\bf 4}, 929 (2000)
  [arXiv:hep-th/0005106].
}

\lref\TakZog{
  P.~G.~Zograf and L.~A.~Takhtajan,
  ``On the Uniformization of Riemann Surfaces and on the Weil-Petersson Metric
  on the Teichm\"uller and Schottky Spaces,''
  Math.USSR-Sb.{\bf 60} (1988), no. 2, 297-313.
}

\lref\McIntyreXS{
  A.~McIntyre and L.~A.~Takhtajan,
  ``Holomorphic factorization of determinants of laplacians on Riemann surfaces
  and a higher genus generalization of Kronecker's first limit formula,''
  arXiv:math/0410294.
}

\lref\LeblancYY{
  Y.~Leblanc,
  ``THE GENUS 2 FREE ENERGY OF THE CLOSED BOSONIC STRING,''
  Phys.\ Rev.\  D {\bf 39}, 3731 (1989).
}

\lref\CoussaertZP{
  O.~Coussaert, M.~Henneaux and P.~van Driel,
  ``The Asymptotic dynamics of three-dimensional Einstein gravity with a
  negative cosmological constant,''
  Class.\ Quant.\ Grav.\  {\bf 12}, 2961 (1995)
  [arXiv:gr-qc/9506019].
}

\lref\TakhtajanMD{
  L.~A.~Takhtajan and L.~P.~Teo,
  ``Quantum Liouville theory in the background field formalism. I: Compact
  Riemann surfaces,''
  Commun.\ Math.\ Phys.\  {\bf 268}, 135 (2006)
  [arXiv:hep-th/0508188].
}

\lref\GaberdielVE{
  M.~R.~Gaberdiel,
  ``Constraints on extremal self-dual CFTs,''
  arXiv:0707.4073 [hep-th].
}

\lref\FriedanUA{
  D.~Friedan and S.~H.~Shenker,
  ``The Analytic Geometry of Two-Dimensional Conformal Field Theory,''
  Nucl.\ Phys.\  B {\bf 281}, 509 (1987).
}

\lref\DijkgraafFQ{
  R.~Dijkgraaf, J.~M.~Maldacena, G.~W.~Moore and E.~P.~Verlinde,
  ``A black hole farey tail,''
  arXiv:hep-th/0005003.
}

\lref\maloneywitten{
 A.~Maloney and E.~Witten, to appear.
}

\lref\YinGV{
  X.~Yin,
  ``Partition Functions of Three-Dimensional Pure Gravity,''
  arXiv:0710.2129 [hep-th].
}

\lref\TakhtajanCC{
  L.~A.~Takhtajan and L.~P.~Teo,
  ``Liouville action and Weil-Petersson metric on deformation spaces, global
  Kleinian reciprocity and holography,''
  Commun.\ Math.\ Phys.\  {\bf 239}, 183 (2003)
  [arXiv:math/0204318].
}

\lref\Hatcher{ W. Jaco, {\sl Lectures on Three-Manifold Topology}, AMS Regional Conference
Series in Mathematics 43. }

\lref\Bogomolny{
  E. Bogomolny, ``Quantum and Arithmetical Chaos",
  arXiv:nlin/0312061 [nlin.CD].
}

\lref\Thurston{
  W. Thurston, {\sl Three-Dimensional Geometry and Topology},
  Princeton University Press, 1997.
}

\lref\Wolpert{
  S. Wolpert, ``Asymptotics of the Spectrum and the Selberg Zeta Function on the
  Space of Riemann surfaces,"
  Commun.\ Math.\ Phys.\ {\bf 112}, 283 (1987).
}

\lref\DHokerZT{
  E.~D'Hoker and D.~H.~Phong,
  ``On Determinants Of Laplacians On Riemann Surfaces,''
  Commun.\ Math.\ Phys.\  {\bf 104}, 537 (1986).
}

\lref\HenningsonGX{
  M.~Henningson and K.~Skenderis,
  ``The holographic Weyl anomaly,''
  JHEP {\bf 9807}, 023 (1998)
  [arXiv:hep-th/9806087].
}

\Title{\vbox{\baselineskip12pt\hbox{} }} {\vbox{\centerline{On Non-handlebody
Instantons in 3D Gravity}}}
\centerline{ Xi Yin }
\smallskip
\centerline{Jefferson Physical Laboratory, Harvard University,
Cambridge, MA 02138} \vskip .6in \centerline{\bf Abstract} { In this
note we describe the contribution from non-handlebody geometries to
the partition function of three-dimensional pure gravity with
negative cosmological constant on a Riemann surface of genus greater
than one, extending previous considerations for handlebodies. }
\vskip .3in

\Date{November 2007}
\listtoc \writetoc \noblackbox

\newsec{Introduction}

The three-dimensional pure quantum gravity with a negative
cosmological constant has been conjectured to be dual to a
holomorphically factorized extremal conformal field theory (ECFT),
of central charge $c=24k$ \WittenKT. While it is not known whether
ECFTs with $k>1$ exist, one may compute its partition function on a
Riemann surface from the gravity path integral, by doing a
perturbative expansion around gravitational instantons and sum over
all instantons. A step toward computing the gravity partition
function was carried out in \YinGV, and it was conjectured that the
classical regularized Einstein-Hilbert action evaluated on a
handlebody hyperbolic instanton agrees\foot{
A (three-dimensional) handlebody is a three manifold homeomorphic to the domain enclosed
by a surface in ${\bf R}^3$.
When the boundary
Riemann surface has genus two, this conjecture was checked to
nontrivial orders near the factorization limit.} with the leading
term in the $1/k$ expansion of the ``fake" CFT partition function,
which captures the part of the full ECFT partition function that
factorizes on Virasoro descendants of the identity operator. It is
then further conjectured in \YinGV\ that the full contribution from
the handlebody instanton is given by the fake CFT partition
function, the latter determined entirely by sphere correlation
functions of Virasoro descendants of the identity.

A priori, one should sum over all hyperbolic three-manifolds $M$ whose conformal boundary is the given Riemann surface
$\Sigma$, of genus $g$. When $g>1$, such manifolds are not all handlebodies. The question remains how to
calculate these non-handlebody contributions. It should be of the form
\eqn\foram{ e^{kS_0+S_1+k^{-1} S_2+\cdots} } where $S_l$ is the $l$-loop contribution
around the instanton background, and depends holomorphically on the moduli of the
Riemann surface. $S_{cl}=-k(S_0+\overline{S_0})$ is the
regularized classical instanton action.
In this note we will describe how to compute $S_{cl}$, and hence
$S_0$. It should have the following properties:

{\bf (1)} $S_{cl}$ is a harmonic function on the moduli space of $\Sigma$, and hence can be written
as $-k(S_0+\overline{S_0})$ for some holomorphic function $S_0$.

{\bf (2)} Let $\Gamma'\subset Sp(2g,{\bf Z})$ be the subgroup of the mapping class
group of $\Sigma$ that extends to $M$ (hence ``preserving" $M$). $e^{S_0}$ transforms
under $\Gamma'$ as a modular form of weight $12$. This is needed to be consistent with
the full partition function transforming as a modular form of weight $12k$.

{\bf (3)} When a handle of $\Sigma$ is pinched, and if $M$ does not fill in the handle,
then $e^{kS_0(M;\Sigma)}$ only contributes to the factorization
on operators in the CFT of dimension $\Delta\geq k$. This is needed to be consistent with the fact that
the handlebody contribution already captures the factorization on operators of $\Delta<k$.\foot{
The first nontrivial primaries in the ECFT have dimension $k+1$, so the factorization on operators
of dimension $\Delta\leq k$ involves Virasoro descendants of the identity only. In \YinGV\ it was found that
the fake CFT partition function, when summed over its modular images, factorizes correctly on states
of dimension $\Delta<k$, at least in the genus two case.
On the other hand, the factorizations on $\Delta=k$ operators
may not be correctly reproduced by the handlebody contributions alone.}

The regularized Einstein-Hilbert action on $M$ has been computed by Krasnov \KrasnovZQ\ when
$M$ is a handlebody, and by Takhtajan and Teo \TakhtajanCC\ for more general hyperbolic manifolds,\foot{
In \TakhtajanCC\ $M$ is required to be of ``Class A", and have in general multiple boundaries, as recalled below.
We are interested in the case where the boundary of $M$ is connected, and $M$ can be lifted to a finite cover
$\tilde M$ which is of Class A.}
and was shown to coincide with a suitably defined Liouville action evaluated at its critical point.
The $S_{cl}$ with the above desired properties, especially property $(1)$, is related to
the Liouville action of \TakhtajanCC\ by a shift, due to conformal anomaly.

In section 2 we will sketch a topological classification of the hyperbolic three-manifold
instantons. Section 3 describes the general strategy in computing the instanton contribution.
In section 4 we consider the factorization
limits of the instanton action.

\newsec{A classification of hyperbolic gravitational instantons}

Consider a hyperbolic three-manifold $M={\bf H}_3/G$, where $G\subset SL(2,{\bf C})$ is a
torsion free
Kleinian group. Suppose that $M$ has a conformal boundary $\Sigma$,
which is a connected Riemann surface of genus $g$. In other words,
$\Sigma=U/G$, where $U={\bf P}^1-\Lambda$ is the domain of
discontinuity for $G$ on the boundary ${\bf P}^1$ of ${\bf H}_3$,
and $\Lambda$ is the set of
limiting points of $G$.

If $G$ is freely generated and purely loxodromic, it is a Schottky group. In this case $M$ is
a handlebody. In general, consider the map
\eqn\maps{i_*:\pi_1(\Sigma)\to \pi_1(M)}
induced by $i:\Sigma=\partial M\hookrightarrow M$.
If $i_*$ is not injective, suppose $\gamma$ is a loop in $\Sigma$ such that $i(\gamma)$
is null-homotopic in $M$. By Dehn's lemma (see for example \Hatcher) there is an embedded disc $D\subset M$ such that
$\partial D=\gamma$. By cutting along $D$, we can reduce $(M,\Sigma)$ to one of the following three
geometries:

\noindent {\bf (i)} $(M',\Sigma')$,
where $M'$ is connected, and $\Sigma'$ has genus $g-1$;

\noindent {\bf (ii)} two disconnected three-manifolds $(M_1',\Sigma'_1)$
and $(M_2',\Sigma_2')$, such that $g_1'+g_2'=g$.

\noindent {\bf (iii)} $(M',\Sigma_1'\sqcup \Sigma_2')$, where $\Sigma_1'$ and $\Sigma_2'$
are the two connected boundary components of $M'$.

Note that in the case (iii) we will be forced to consider manifolds with multiple boundaries. Such
gravitational instantons are rather pathological, as will be discussed later.
By repeating such surgeries, we can reduce $M$ to one or
several disconnected three-manifolds whose boundaries are $\pi_1$-injective.
We will call the hyperbolic manifold $M$ with a $\pi_1$-injective
connected conformal boundary $\Sigma$ a ``tight" manifold. A simple class of tight manifolds are given topologically by
twisted $I$-bundles over an unoriented surface $S$, namely
$I\to M\to S$, such that $\Sigma$ is a two-fold covering of $S$.
These are in fact all tight manifolds with the property that $i_*\pi_1(\Sigma)$ is a finite index
subgroup of $\pi_1(M)$ (and the index is 2).
On the other hand, there are also tight manifolds with $[\pi_1(M):i_*\pi_1(\Sigma)]=\infty$.\foot{
I'm grateful to C. McMullen for explaining to me such examples.}

\vskip 0.5cm
\centerline{\vbox{\centerline{ \hbox{\vbox{\offinterlineskip
\halign{&\strut\hfill#\hfill\hskip0.4cm&\strut\hskip0.2cm \hfill #\hfill\hskip0.2cm
&\strut\hskip0.2cm \hfill #\hfill\hskip0.2cm\cr
\epsfysize=1.3in \epsfbox{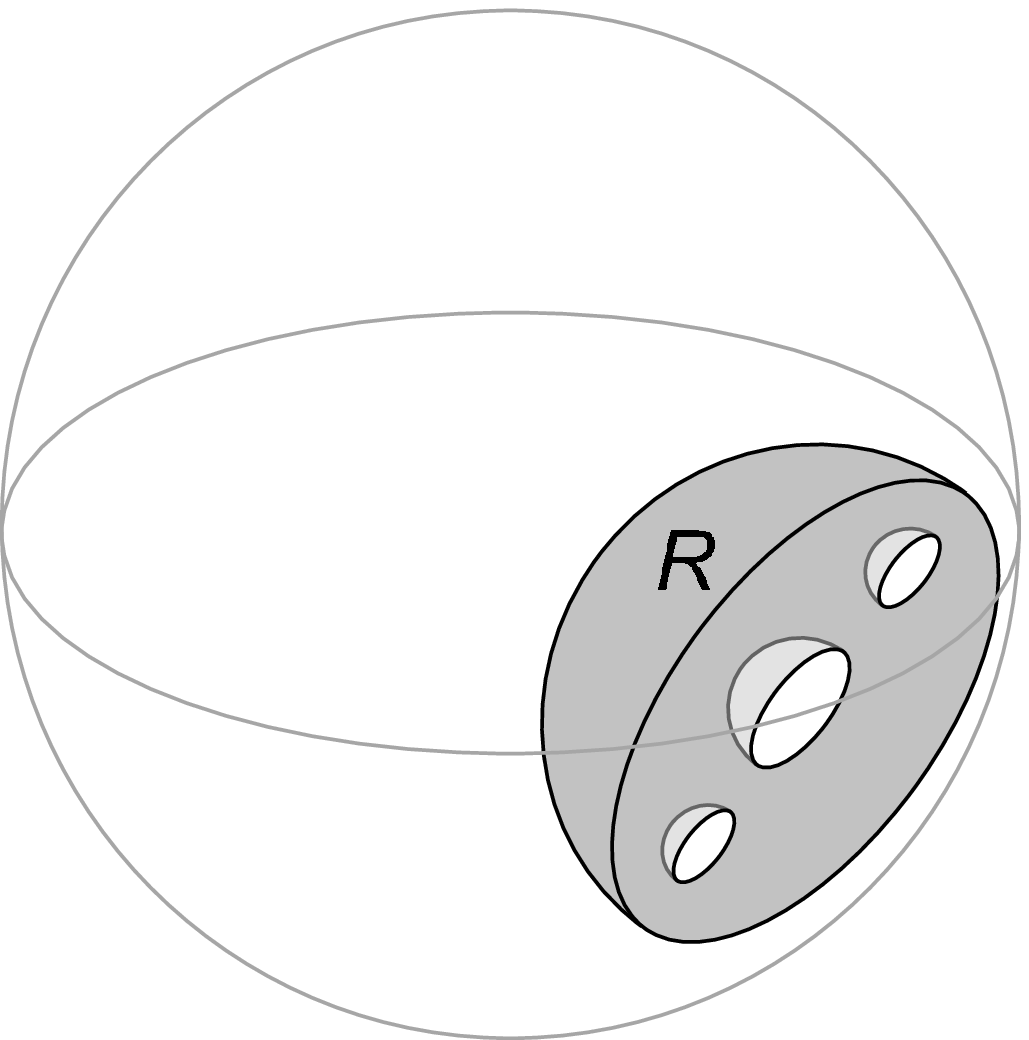}& \epsfysize=1.3in \epsfbox{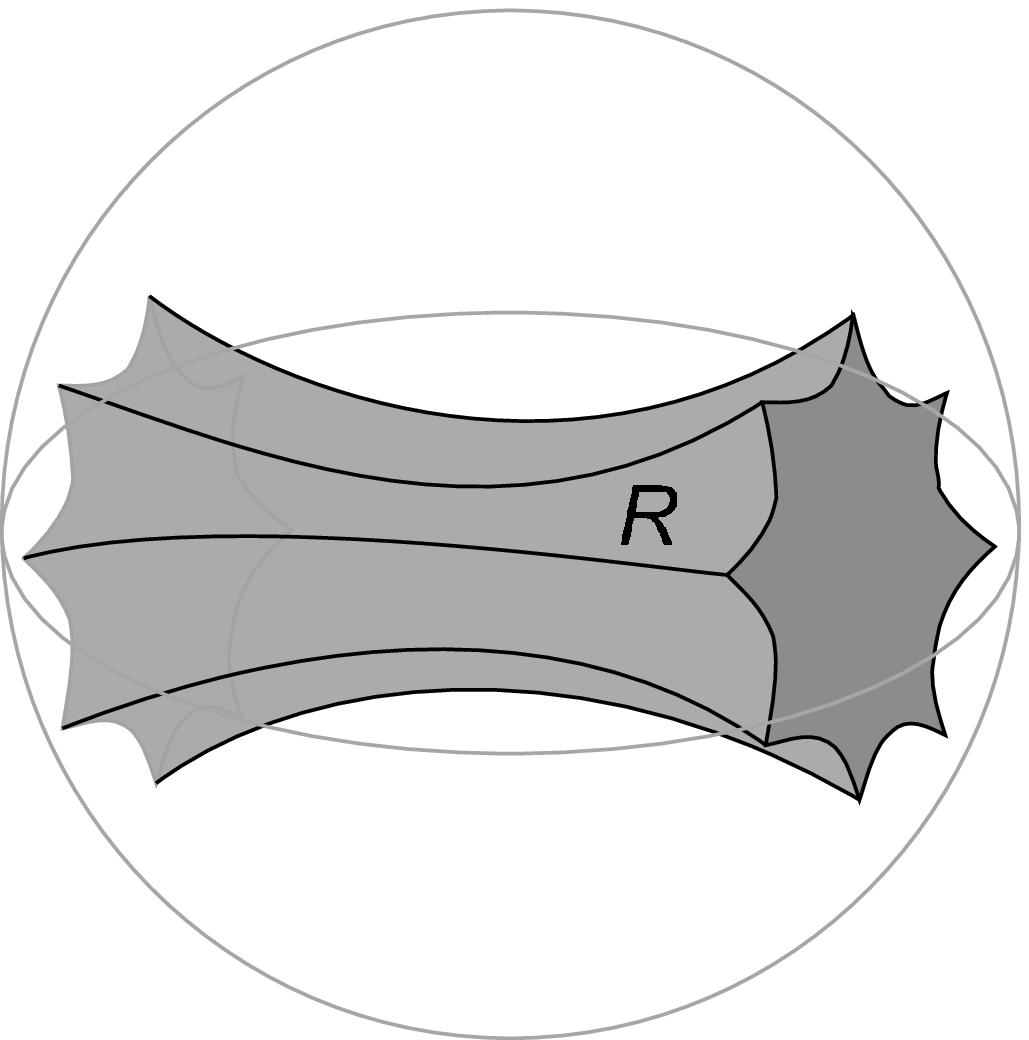} &
\epsfysize=1.3in \epsfbox{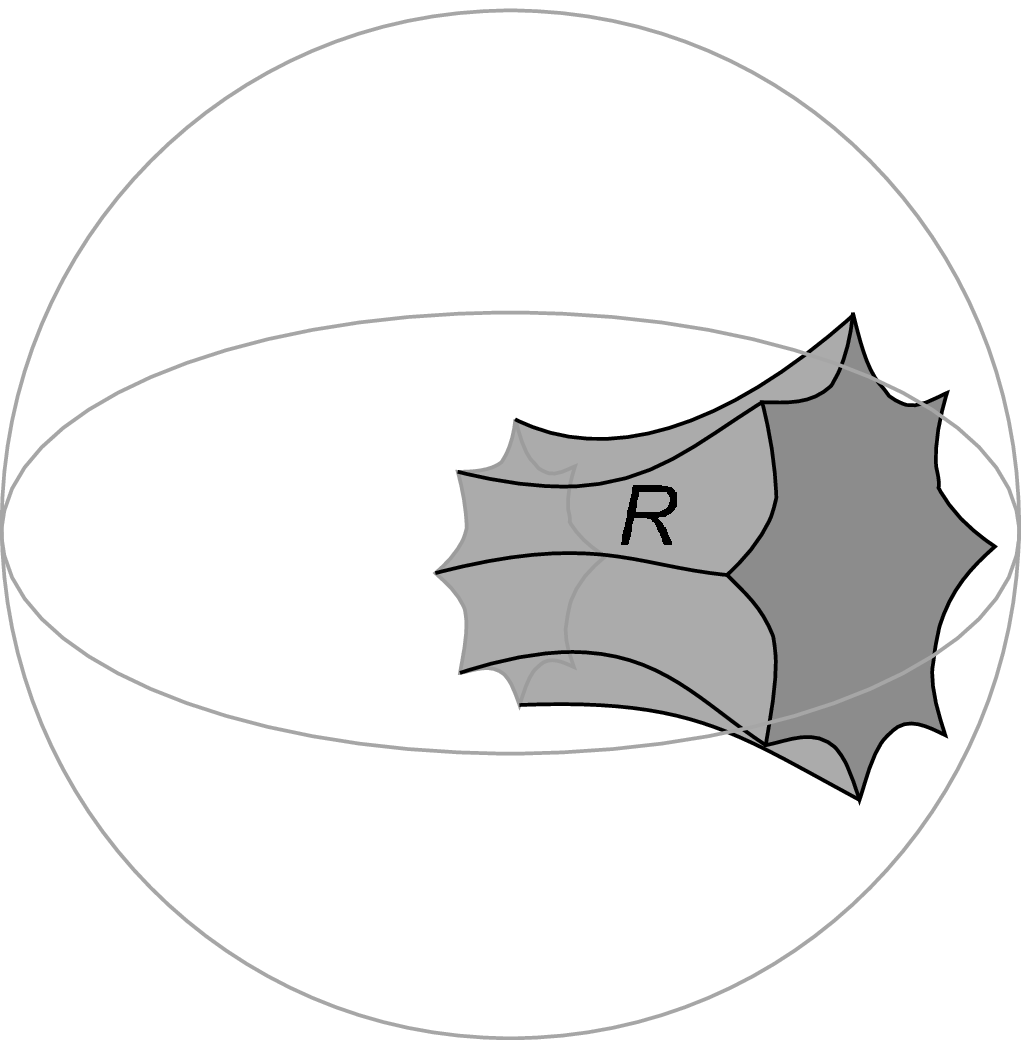}  \cr (a) & (b) & (c) \cr }}}}
{{\bf Figure 1.} A typical fundamental domain $R$ of (a) a
handlebody, (b) a class A manifold with two boundaries, and (c) a
non-handlebody with one boundary, in the hyperbolic 3-space ${\bf
H}_3$.} }} \vskip 0.5cm

A fundamental domain for $M={\bf H}_3/G$ in ${\bf H}_3$ is of the form $(R,F)$, where $R$ is a fundamental
domain of $G$ in ${\bf H}_3$, and $F=G\cap U$ a fundamental domain for $\Sigma=U/G$.
In general, $R$ can be described as a cell complex, with $3,2,1,0$-cells, corresponding to the bulk of $R$,
its faces, edges, and corners. $G$ is called a
``Class A" Kleinian group if one can choose the fundamental domain
$R$ to have no $0$-cells in the bulk of ${\bf H}_3$. General
non-handlebody class A manifolds $M$ will have multiple boundary components, $\Sigma_1,\cdots,
\Sigma_n$. We shall consider the case when $(M,\Sigma)$ can be lifted to a finite covering space
$(\tilde M,\Sigma_1\sqcup\cdots\sqcup\Sigma_n)$, such that $\tilde M$ is of class A.

\newsec{The holomorphically factorized classical action of $M={\bf H}_3/G$}

The regularized Einstein-Hilbert action of \refs{\HenningsonGX,\KrasnovZQ} takes the
form \eqn\regeh{ S_{EH}(M;\Sigma) = {4k\over\pi} \lim_{\epsilon\to
0} \left(V_\epsilon - {1\over 2} A_\epsilon +2\pi(2g-2)\ln\epsilon
\right) } Here $V_\epsilon$ and $A_\epsilon$ are the volume of the
bulk hyperbolic three-manifold and the area of the boundary cutoff
surface, respectively; $\epsilon$ is the cutoff parameter. The
cutoff surface is chosen so that its induced metric has constant
curvature $-1/\epsilon^2$. The $\ln\epsilon$ divergence in the
Einstein-Hilbert action with boundary term is related to the
conformal anomaly in the boundary CFT.

The main result of \TakhtajanCC\ is that, if $(\tilde M,\Sigma_1\sqcup\cdots\sqcup\Sigma_n)$
is of class A, then the regularized Einstein-Hilbert
action on $\tilde M$ is related to the classical Liouville action evaluated at its critical point,
$S_L(\tilde M,\Sigma_1\sqcup\cdots\sqcup\Sigma_n)$, by
\eqn\liuvv{ S_{EH}(\tilde M,\Sigma_1\sqcup\cdots\sqcup\Sigma_n) = -k\left[S_L(\tilde M,\Sigma_1\sqcup\cdots\sqcup\Sigma_n)
+\sum_{i=1}^n(2g_i-2)\cdot const\right].  }
where $g_i$ is the genus of $\Sigma_i$.
We refer to \refs{\KrasnovZQ,\TakhtajanCC} for the precise definition of $S_L$ (which, importantly, depends
not only on $\Sigma$ but on the
Kleinian group $G$ as well).\foot{Our convention for $S_L$ differs from that
of \TakhtajanCC\ by a factor of $\pi$.} An important property is that
$S_L$ is a Kahler potential for the Weil-Petersson metric on the Teichm\"uller space of
$\Sigma_1\sqcup\cdots\sqcup\Sigma_n$. More generally, if $M$ is not in class A but
can be lifted to its $n$-fold covering space $\tilde M$ which is in class A, then the regularized
Einstein-Hilbert action on $M$ is given by
\eqn\abss{ S_L(M;\Sigma) = {1\over n}S_L(\tilde M;\Sigma,\cdots,\Sigma) }
Clearly, $S_L(M;\Sigma)$ will also be a Kahler potential for the Weil-Petersson metric
on the Teichm\"uller space of $\Sigma$.
Consequently, if $M_1, M_2$ have the same conformal boundary
$\Sigma$, then $S_L(M_1;\Sigma)-S_L(M_2;\Sigma)$ is a harmonic function on the Teichm\"uller space of
$\Sigma$, i.e.
$\exp(S_L(M_1;\Sigma)-S_L(M_2;\Sigma))$ is holomorphically factorized.

We define the ``holomorphically factorized" classical action
$S_{cl}=-kS_0$ by \eqn\szero{\eqalign{ S_0(M;\Sigma) +
\overline{S_0(M;\Sigma)}& = S_L(M;\Sigma)+12 \ln {\det'\Delta\over
\det{\rm Im}\Omega} \cr &= S_L(M;\Sigma)+12 \ln
{\zeta_\Sigma'(1)\over \det{\rm Im}\Omega} + (2g-2)c_0 } } where
$\zeta_\Sigma(s)$ is the Selberg zeta function for the Riemann
surface $\Sigma$ \DHokerZT, and $c_0$ is a constant. By Zograf's
factorization formula for $\det'\Delta$ \McIntyreXS, the RHS of
\szero\ is harmonic when $M$ is a handlebody; by the above argument,
this must also be the case for all $M$ whose boundary is $\Sigma$.
\szero\ still leaves the ambiguity of adding an imaginary constant
to $S_0(M;\Sigma)$, which may depend on the topology of $M$; this
corresponds to the overall phase of the contribution $e^{kS_0}$ to
the holomorphic partition function. A natural choice of the phase is
such that $e^{S_0}$ is real when ${\rm Re}\Omega=0$. This is
consistent with the factorization of the partition function. This
still leaves an overall sign ambiguity for $e^{S_0}$. The sign may
potentially be different for distinct topologies.

The full quantum holomorphic partition function on $M$,
$Z_k(M;\Sigma)$, should be a weight $12k$ holomorphic modular form
under $\Gamma_G\subset Sp(2g,{\bf Z})$, the subgroup of the mapping
class group of $\Sigma$ that leaves $M$ invariant, defined on the
Teichm\"uller space of $\Sigma$.\foot{When $M$ is a handlebody,
$Z_k(M;\Sigma)$ is simply invariant under $\Gamma_\infty$. We are
working in the convention that the partition function of a chiral
boson on $\Sigma$ is normalized to 1; in other words, $Z$ is the
partition function of the holomorphic CFT divided by that of $24k$
chiral bosons.} It takes the form \eqn\caba{ Z_k(M;\Sigma) =
\exp\left[ k S_0(M;\Sigma) + S_1(M;\Sigma) + {1\over k}
S_2(M;\Sigma)+\cdots \right] } where $S_1,S_2,\cdots$ are loop
corrections, suppressed by powers of $1/k$. In \szero, $\det'\Delta$
is modular invariant, and $S_L(M;\Sigma)$ is invariant under $G$.
Due to the $\det{\rm Im}\Omega$ factor, $|e^{kS_0}|^2$ transforms
under $\Gamma_G$ with holomorphic and anti-holomorphic weight $12k$,
as expected.

Suppose $(M,\Sigma)$ can be reduced to $(M',\Sigma')$ by cutting
along an embedded disc $(D,\partial D)$. For a general holomorphic
CFT of central charge $c=24k$, the partition function on $\Sigma$
and $\Sigma'$ are related by \eqn\zzf{ Z(\check\Sigma) =
G({\Sigma',z_1,z_2};q)^k \sum_i q^{\Delta_i}\langle {\cal
A}_i(z_1){\cal A}_i(z_2)\rangle_{\check\Sigma'} } where $\Sigma$ is
obtained from $\Sigma'$ by gluing a handle of modulus $q$ to
$z_1,z_2$. The notation $\check\Sigma,\check\Sigma'$ indicates a
compatible choice of basis 1-cycles on the Riemann surfaces, as the
partition functions are modular forms of nonzero weight. $G$ is a
universal holomorphic correction factor that depends only on the
gluing procedure, with the property $G(\Sigma',z_1,z_2;q=0)=1$. The
conjecture of \YinGV\ is that we can compute the gravity partition
function of $\Sigma$ from that of $\Sigma'$ by \eqn\zfakedef{
Z_k(M;\Sigma) = G(\Sigma',z_1,z_2;q)^k \sum_{{\cal A}_i\in Vir(k)}
q^{\Delta_i} \langle {\cal A}_i(z_1){\cal
A}_i(z_2)\rangle_{fake;M',\Sigma'} } where the sum is only over
Virasoro descendants of the identity (denoted by $Vir(k)$). On the
RHS, the ``fake" two-point function of ${\cal A}_i\in Vir(k)$ on
$\Sigma'$ is completely determined by $Z_k(M';\Sigma')$, since all
correlators of the stress tensor on $\Sigma'$ can be obtained by
taking derivatives of $Z_k(M';\Sigma')$ with respect to the complex
moduli.

\vskip 0.5cm
\centerline{\vbox{\centerline{ \hbox{\vbox{\offinterlineskip
\halign{&#&\strut\hskip0.2cm \hfill #\hfill\hskip0.2cm\cr
\epsfysize=.9in \epsfbox{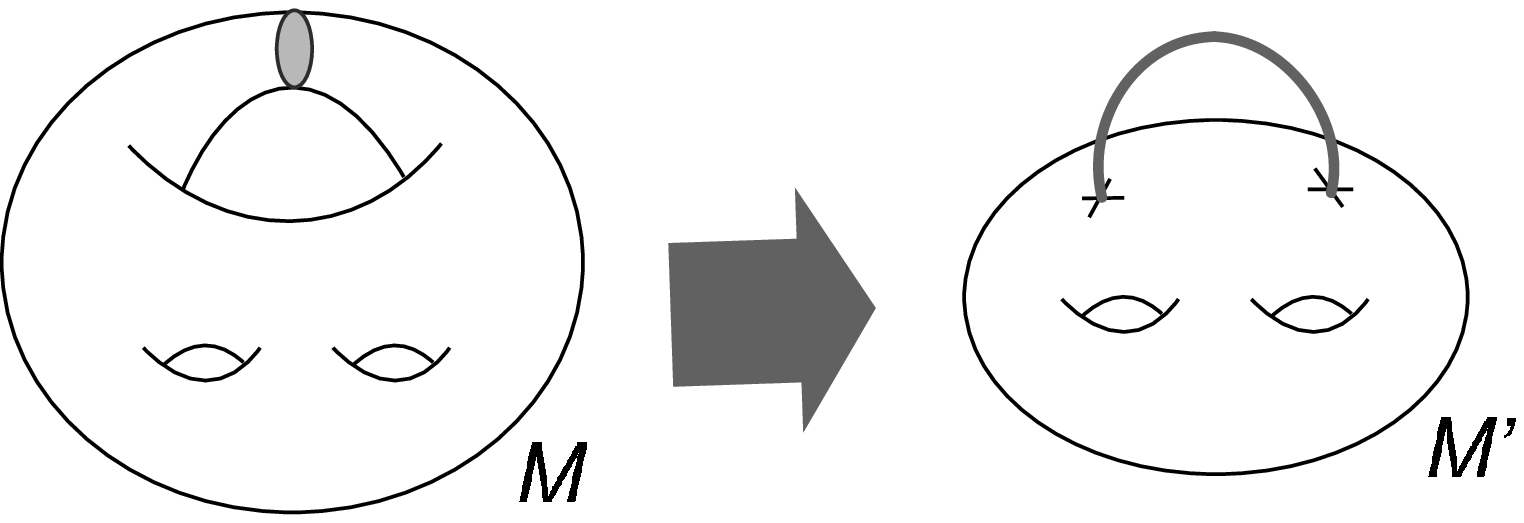}  \cr }}}}
\centerline
{{\bf Figure 2.} Reducing $M$ to $M'$ along a filled handle.}
}} \vskip 0.5cm

When $M'$ has two connected components $M'_1$ and $M'_2$, with
conformal boundary $\Sigma_1'$ and $\Sigma_2'$, the gravity
partition functions can be similarly related as \eqn\fakedisc{
Z_k(M,\Sigma) = G(\Sigma_1',z_1;\Sigma_2',z_2;\epsilon)^k
\sum_{{\cal A}_i\in Vir(k)} \epsilon^{\Delta_i} \langle {\cal
A}_i(z_1) \rangle_{fake;M'_1,\Sigma'_1} \langle {\cal A}_i(z_2)
\rangle_{fake;M'_2,\Sigma'_2} } where $\Sigma$ is obtained by sewing
$\Sigma_1'$ and $\Sigma_2'$ together along a tube of modulus
$\epsilon$, attached to the points $z_1\in \Sigma_1'$ and $z_2\in
\Sigma_2'$. $G(\Sigma_1',z_1;\Sigma_2',z_2;\epsilon)$ is the
appropriate holomorphic correction factor in factorizing a ($c=24$)
CFT partition function on $\Sigma$ into the one-point functions on
$\Sigma_1'$ and $\Sigma_2'$, with the property
$G(\Sigma_1',z_1;\Sigma_2',z_2;\epsilon=0)=1$.\foot{When $\Sigma_1'$
and $\Sigma_2'$ are of genus one,
$G(\Sigma_1',z_1;\Sigma_2',z_2;\epsilon)$ is related to the
holomorphic correction factor of \refs{\Tuite,\YinGV} by a
normalization factor
$\chi_{10}(\Omega)/(\epsilon^2\Delta(\tau_1)\Delta(\tau_2))$, due to
our different convention of the genus $g$ partition function.}

\vskip 0.5cm
\centerline{\vbox{\centerline{ \hbox{\vbox{\offinterlineskip
\halign{&#&\strut\hskip0.2cm \hfill #\hfill\hskip0.2cm\cr
\epsfysize=.6in \epsfbox{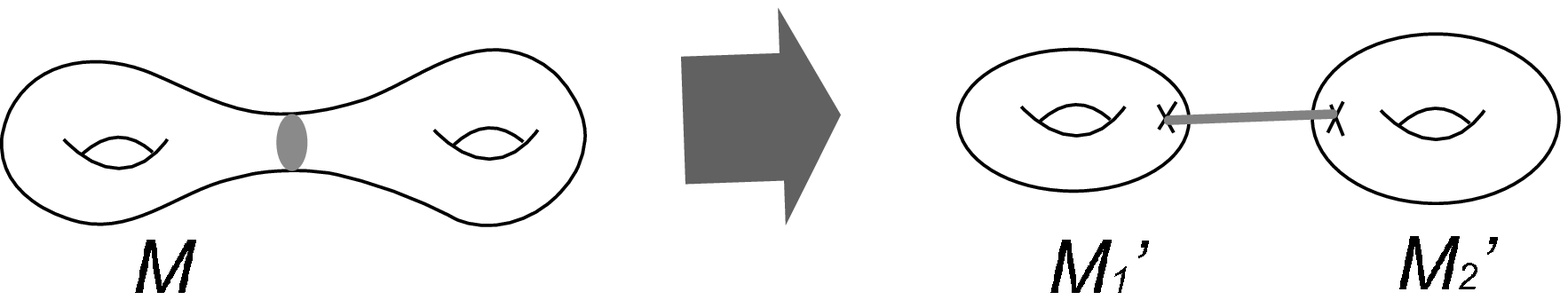}  \cr }}}}
\centerline
{{\bf Figure 3.} Reducing $M$ to $M'_1\sqcup M_2'$ along a filled tube.}
}} \vskip 0.5cm

Finally, when $M'$ is connected but have two boundary components
$\Sigma_1'$ and $\Sigma_2'$, the contribution from $M$ and $M'$
should be related by \eqn\zfakedef{ Z_k(M;\Sigma) =
G(\Sigma_1',z_1;\Sigma_2',z_2;\epsilon)^k \sum_{{\cal A}_i\in
Vir(k)} \epsilon^{\Delta_i} {\cal D}^{{\cal A}_i(z_1)}_{\Sigma_1'}
{\cal D}^{{\cal A}_i(z_2)}_{\Sigma_2'} Z_k({M';\Sigma_1'\sqcup
\Sigma_2'})
} Here ${\cal D}^{{\cal A}(z)}_{\Sigma}$ is a differential operator
in the moduli of $\Sigma$, defined by the property ${\cal D}^{{\cal
A}(z)}_{\Sigma}Z(\Sigma) = \langle {\cal A}(z)\rangle_{\Sigma}$,
where $Z(\Sigma)$ and $\langle {\cal A}(z)\rangle_\Sigma$ are the
partition function and one-point function of a general $c=24k$ CFT
on $\Sigma$, ${\cal A}\in Vir(k)$.

\vskip 0.5cm
\centerline{\vbox{\centerline{ \hbox{\vbox{\offinterlineskip
\halign{&#&\strut\hskip0.2cm \hfill #\hfill\hskip0.2cm\cr
\epsfysize=.8in \epsfbox{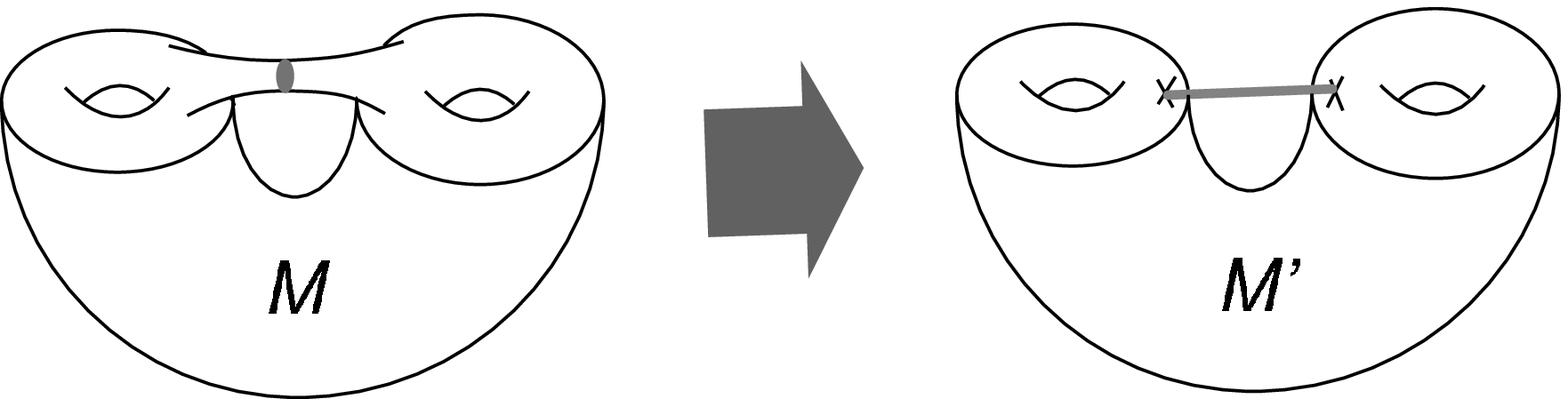}  \cr }}}}
\centerline
{{\bf Figure 4.} Reducing $M$ to $M'$ with two boundary components.}
}} \vskip 0.5cm

The last case is however puzzling from the dual CFT perspective, as
it appears to spoil the factorization of the partition function on
$\Sigma$ into the product of the partition functions on $\Sigma_1'$
and $\Sigma_2'$ in the pinching limit, barring miraculous
cancelations. There are two possible interpretations: (1) the dual
CFT does not exist, due to the failure of the factorization of the
gravity partition function; or (2) gravitational instantons that
lead to connected $M'$'s with multiple boundary components under the
cutting surgery (Figure 4) should be excluded from the gravity path
integral. Note that since $M$ is a hyperbolic manifold, it is
atoroidal, which implies that $\Sigma_1'$ and $\Sigma_2'$ must have
genus $g_1', g_2'>1$.\foot{ An explicit example of such $M$ is
obtained topologically by attaching a solid handle to the two sides
of $\Sigma'\times I$, where $\Sigma'$ is a genus $g>1$ surface. Now
$M$ has a genus $2g$ boundary, and admits a hyperbolic metric. The
corresponding Kleinian group $G$ is a free product of a
(quasi-)Fuchsian group with ${\bf Z}$, the latter generated by a
loxodromic element of $SL(2,{\bf C})$ of sufficiently large
multiplier.} So the potential failure of the factorization of the
partition function on $\Sigma$ can only show up at genus $g\geq 4$.
Also note that $[\pi_1(M):i_*\pi_1(\Sigma)]=\infty$ in this case.
From now on we will adopt the second interpretation above, and
exclude these pathological gravitational instantons. This may seem
rather ad hoc from the perspective of the gravity path integral; on
the other hand, it leads to dual CFT partition functions with
consistent factorization property, and one may be able to extract
CFT correlation functions from them.

In general, the above conjectured relations between $Z_k(M,\Sigma)$
for different pairs $(M,\Sigma)$ related by surgeries encode
nontrivial relations between fake Virasoro correlators and the
Liouville action $S_L(M;\Sigma)$, generalizing the conjectures of
\YinGV\ for handlebodies.

We know $S_0$ explicitly in two special classes of examples. When $M$ is a handlebody,
as explained in \YinGV,
\eqn\sosah{ S_0(M;\Sigma) = 12\sum_{\gamma~ prim.} \sum_{m=1}^\infty \ln(1-q_\gamma^m) }
where the sum runs through all primitive conjugacy classes of the Schottky group $G$,
and $q_\gamma$ is the multiplier of $\gamma\in SL(2,{\bf C})$, with $|q_\gamma|<1$.

\vskip 0.5cm
\centerline{\vbox{\centerline{ \hbox{\vbox{\offinterlineskip
\halign{&#&\strut\hskip0.2cm \hfill #\hfill\hskip0.2cm\cr
\epsfysize=1.9in \epsfbox{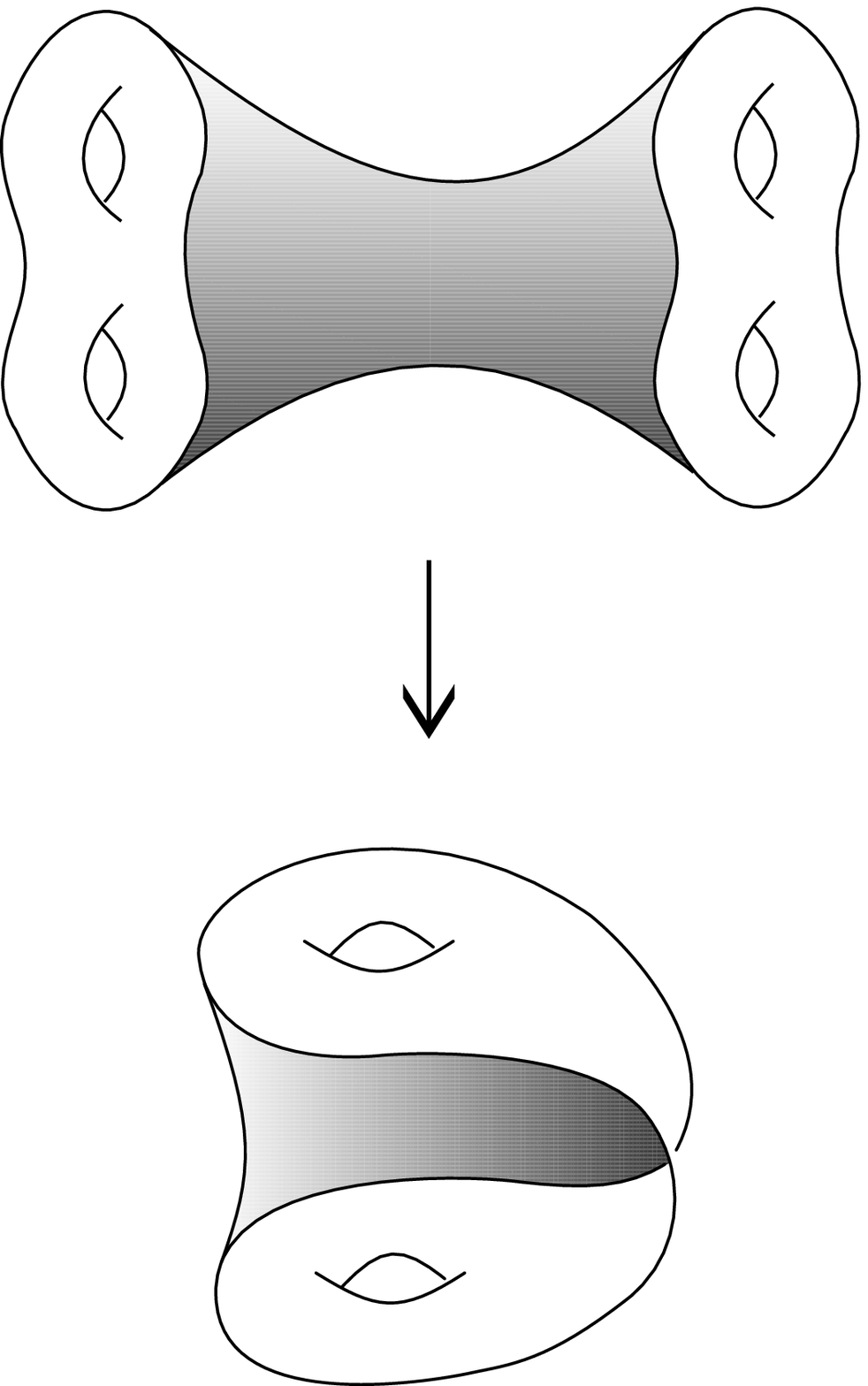}  \cr }}}}
\centerline
{{\bf Figure 5.} The twisted $I$-bundle as a ${\bf Z}_2$ quotient.}
}} \vskip 0.5cm

When $M$ is {\sl topologically} a twisted $I$-bundle $I\to M\to S$,
we can lift $(M;\Sigma)$ to its double cover $(\tilde
M;\Sigma\sqcup\Sigma)$, such that $M=\tilde M/\iota$ for an
involution $\iota$ on $\tilde M$. When $\Sigma=\overline{\Sigma}$,
$\tilde M$ is the quotient of ${\bf H}_3$ by a Fuchsian group
$\tilde G$. In this case we can explicitly write the hyperbolic
metric on $\tilde M$ as \eqn\hysaa{ ds^2 = dr^2 + \cosh^2r
ds_\Sigma^2, } where $ds_\Sigma^2$ is a hyperbolic metric on
$\Sigma$. $\iota$ acts as $r\to -r$ together with an orientation
reversing, fixed-point free involution on $\Sigma$. For example,
suppose $\Sigma$ has genus two, with period matrix $\Omega =
\pmatrix{\rho & \nu \cr \nu & \sigma }$. Then $\Sigma$ admits such
an involution at the real locus of its moduli space,
$\rho=-\bar\sigma$, $\nu=i\nu_2$, $\nu_2\in{\bf R}$. At a generic
point on the moduli space, the metric on $\tilde M$ does not take
the form \hysaa, and $\tilde G$ will be a quasi-Fuchsian group
instead of a Fuchsian group (it fixes a Jordan curve on $\partial
{\bf H}_3={\bf P}^1$, rather than the equator of the ${\bf P}^1$).

In the case $\Sigma=\bar\Sigma$ and $\tilde G$ is a Fuchsian group, the Liouville action simply
evaluates to $S_L(\tilde M;\Sigma,\Sigma)=2S_L(M;\Sigma)=2c(2g-2)$
for some constant $c$ \refs{\TakZog,\TakhtajanCC}. In other words,
\eqn\sozz{ 2{\rm Re}S_0(M;\Sigma=\overline{\Sigma})=c(2g-2)+12 \ln
{\zeta_\Sigma'(1)\over \det{\rm Im}\Omega_\Sigma} }
$\zeta_\Sigma(s)$ can be defined as \eqn\sficc{ \zeta_\Sigma(s) =
\prod_{\Upsilon ~prim.} \prod_{m=0}^\infty (1-q_\Upsilon^{m+s}), }
where the first product is over all primitive conjugacy classes of
the Fuchsian group of $\Sigma$, and $q_\Upsilon$ is the multiplier
of $\Upsilon\in SL(2,{\bf R})$, $q_\Upsilon<1$. $\Upsilon$ also
corresponds to a primitive geodesic on the surface $\Sigma$ equipped
with hyperbolic metric, and $q_\Upsilon=e^{-l(\Upsilon)}$ where
$l(\Upsilon)$ is the length of the geodesic. \sozz\ could be used to
determine the harmonic function ${\rm Re}S_0(M;\Sigma)$ on the
entire Teichm\"uller space of $\Sigma$.

Now that we know how to compute $S_0$, at least in principle, a
remaining question is how to compute the $1/k$ corrections $S_1,
S_2, \cdots$ for tight manifolds $(M,\Sigma)$. Once these are known,
$Z_k(M,\Sigma)$ for the tight manifolds will be determined, and it
can be used to determine the partition functions of all $(M,\Sigma)$
by the sewing procedure described earlier. It would also be nice to
have a formula analogous to \sosah\ for all $M$ (non-handlebodies).

\newsec{Factorization}

\subsec{Degenerating limits of Selberg zeta function}

Starting with a Riemann surface $\Sigma$ of genus $g$, let us
consider the limit where a handle is pinched, and $\Sigma$ is
reduced to a Riemann surface $\Sigma'$ of genus $g-1$. In order to
examine the behavior of Selberg zeta function on $\Sigma$ in this
limit, we will assume that $\Sigma$ is equipped with a hyperbolic
metric, and let the length of the short geodesic around the pinched
handle be $2\pi l$. Along the pinched handle, the metric can be
approximated by the hyperbolic metric on an infinite tube,
\eqn\tuba{ d\chi^2 +l^2{\cosh^2\chi} d\phi^2 = {|dw|^2\over
\sin^2({\rm Re}\,w)} } where \eqn\wexp{ w = -i\ln {\sinh
\chi+i\over\cosh \chi}+il{\phi},~~~~ {\rm Re}\,w\in (0,\pi) } The
modulus of the tube, $\tau$, is related to the length of the short
geodesic by $\tau_2=\pi/(2\pi l)=1/(2l)$. $dw$ approximates a
holomorphic 1-form on $\Sigma$, restricted to the tube. The period
matrix of $\Sigma$ takes the form \eqn\aome{ \Omega_\Sigma \to
\pmatrix{ \tau & * \cr * & \Omega_\Sigma' } } in the pinching limit,
where $*$ stands for finite entries.

We can write the Selberg zeta function $\zeta_\Sigma(s)$ as
\eqn\contrsh{ \zeta_\Sigma(s)=\tilde f_\Sigma(s)\prod_{m=0}^\infty
(1-e^{-2\pi l(m+s)})^2, } where we singled out the contributions
from the short geodesics (counted with both orientations). $\tilde
f_\Sigma(s)$ denotes the contribution from all other closed
geodesics on $\Sigma$. $\tilde f_\Sigma(s)$, just like
$\zeta_\Sigma(s)$, has a simple zero at $s=1$. This can be
understood from the fact that the density of closed geodesics of
length $L$ grows as $\rho(L)\sim e^{-L}/L$ \Bogomolny. Therefore we
have \eqn\dgoa{ \zeta_\Sigma'(1) = \tilde
f_\Sigma'(1)\prod_{m=1}^\infty (1-e^{-2\pi ml})^2.  } Now ${\tilde
f_\Sigma}'(1)$ is finite in the $l\to 0$ limit, since it does not
involve the contribution from the short geodesic of length $2\pi l$.

Using the modular transformation of the Dedekind eta function
\eqn\etafuf{ \prod_{m=1}^\infty (1-e^{-2\pi ml}) = e^{\pi
l\over 12}\eta(il) = {e^{\pi l\over 12}\over \sqrt{l}}\eta(i/l) \sim
{e^{-{\pi\over 12l}}\over\sqrt{l}},~~~~~(l\to 0) } and $\det{\rm
Im}\Omega_\Sigma\sim \tau_2\sim {1\over 2l}$, we find \eqn\asrft{
\ln{\zeta_\Sigma'(1)\over\det {\rm Im}\Omega_\Sigma} \sim
{-{\pi\over 3}\tau_2}+ finite } in the pinching limit. This result
is also known from \Wolpert.

Let us now consider the limit in which $\Sigma$ is pinched at a tube
connecting two components $\Sigma_1'$ and $\Sigma_2'$, of genus
$g_1'$ and $g_2'$, $g_1'+g_2'=g$. Along the tube, the metric is
again approximated by \tuba; the modulus of the tube is related to
$l$ in the same way as before. A difference is that, in the
separating degeneration limit, \eqn\omegasep{ \Omega_\Sigma \to
\pmatrix{ \Omega_{\Sigma_1'} & 0 \cr 0 & \Omega_{\Sigma_2'}, } } and
in particular $\det{\rm Im}\Omega_\Sigma$ is non-degenerate. The
Selberg zeta function takes the form \eqn\sellm{ \zeta_\Sigma(s) \to
f_1(s) f_2(s) \prod_{m=0}^\infty (1-e^{-2\pi l(m+s)})^2 } where
$f_1(s)$ and $f_2(s)$ involve closed geodesics on $\Sigma_1'$ and
$\Sigma_2'$, respectively. At the separating degeneration, each
$f_i(s)$ has simple zero at $s=1$, since the number of geodesics of
length $\sim L$ on each punctured Riemann surface $\Sigma_i'$ grows
like $\rho(L)\sim e^L/L$. In particular,
$\partial_s|_{s=1}(f_1(s)f_2(s))\to 0$ in the $l\to 0$ limit. In
order to extract the $l$ dependence, we make use of the estimates of
\Wolpert \eqn\aaasel{ \zeta_\Sigma'(1) \sim
\lambda_1\prod_{m=1}^\infty (1-e^{-2\pi ml})^2\times finite,~~~~l\to
0 } where $\lambda_1$ is the smallest nonzero eigenvalue of the
Laplacian $\Delta$ on $\Sigma$ (all other nonzero eigenvalues of
$\Delta$ are of order 1 in the degeneration limit). It is easy to
see that $\lambda_1\sim l$, and hence \asrft\ still holds in the
separating degeneration limit.

\subsec{$S_L$ in the factorization limits}

Now let us consider the Liouville action $S_L$ in the factorization
limits. A special case is when $M$ is a twisted $I$-bundle, and the
complex structure of $\Sigma$ is such that it admits an orientation
reversing ${\bf Z}_2$ involution. As discussed in section 3, $S_L$
takes constant value along this real locus of the moduli space of
$\Sigma$. It then immediately follows from \szero\ that when a
handle or tube of $\Sigma$ is pinched (along the real locus), the
contribution from $M$ to the partition function behaves like
\eqn\statbeh{ e^{kS_0(M;\Sigma)} \sim q^k f(M',\Sigma') + {\cal
O}(q^{k+1}) } where $q=e^{2\pi i\tau}$ is the pinching modulus
parameter, and $f$ is a generic function that depends on the pinched
geometry. This means that $M$ can only contribute to the
factorization on states of dimension $\Delta\geq k$ along any tube
of the Riemann surface.

More generally, the Liouville action $S_L(M;\Sigma)$ in \szero\ is
bounded when a cycle corresponding to an element $\gamma$ of $G$ is
pinched, or equivalently, when the pinched loop is not contractible
in $M$. In fact, the thin tube that is being pinched is formed by
gluing a thin strip of the fundamental domain $F$ of $\Sigma$ on the
${\bf P}^1$ by identifying the two sides by $\gamma$, and
$\gamma'(z)$ approaches 1 in the pinching limit along the strip. One
may worry about the potential divergence in $S_L$ due to the
singular behavior of the Liouville field $\phi$ near the pinching point. To see this,
let us represent the thin strip as the domain between two circles in
the complex $z$-plane, both centered on the real axis and touching
say at $z=0$ (Figure 6). Near the pinching point, the Liouville
field $\phi$ is approximately given by \eqn\phis{ \phi \simeq -\ln
({\rm Im}z)^2, } and so that $|\partial_z \phi|^2\simeq e^\phi$. It
follows that in the Liouville action $S_L$, both the integral of the
Liouville Lagrangian over the bulk of the fundamental domain $F$, as
well as the boundary integrals, are finite. The singular behavior of
$S_0$ \szero\ then entirely comes from the term $12\ln
(\zeta'_\Sigma(1)/\det {\rm Im}\Omega)$, as analyzed earlier.

\vskip 0.5cm
\centerline{\vbox{\centerline{ \hbox{\vbox{\offinterlineskip
\halign{&#&\strut\hskip0.2cm \hfill #\hfill\hskip0.2cm\cr
\epsfysize=1.9in \epsfbox{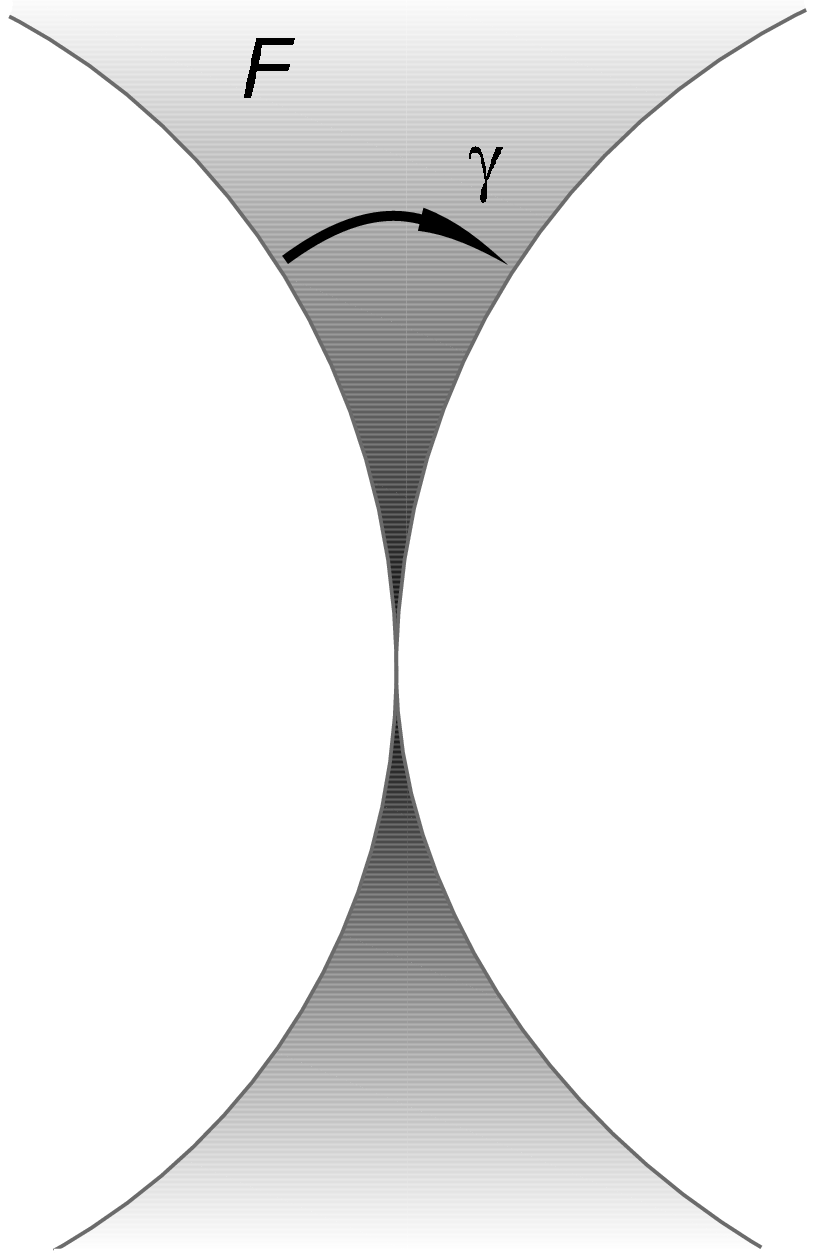}  \cr }}}}
\centerline
{{\bf Figure 6.} The strip in the fundamental domain of $\Sigma$ corresponding to
the pinched tube.}
}} \vskip 0.5cm

We conclude that when a loop $\gamma$ of $\Sigma$ is pinched, if
$\gamma$ is incontractible in $M$, then $e^{kS_0(M;\Sigma)}$ can
only contribute to the factorization of the partition function on
states of dimension $\geq k$.

Note that if $\gamma$ were contractible in $M$, the pinching limit
would correspond to shrinking a pair of circles whose interiors are
excluded from $F$, rather than having two circles touching one
another. In this case the Liouville action $S_L$ will generically
diverge. For example, suppose a circle $C:|z|=r_0$ is identified
with $C':|z-z_0|=r_0$ via the action \eqn\gamact{ \gamma(z) =
{-e^{2i\theta_0}r_0^2\over z}+z_0 } We have $|\gamma'(z)|=1$ along
$C$. The Liouville action receives the contribution \refs{\McIntyreXS,\TakhtajanCC}
\eqn\phialiu{{1\over\pi} \int_F d^2z (|\partial_z\phi|^2+e^\phi) -
{i\over\pi}\oint_C \phi d\ln{\gamma'(z)} +4 \ln |c(\gamma)|^2 } from
the domain near $C$ and $C'$. Here $c(\gamma)=c=e^{-i\theta_0}/r_0$
for $\gamma=\pmatrix{a & b\cr c & d} \in SL(2,{\bf C})$. Near $C$,
in the $r_0\to 0$ limit, the Liouville field behaves as
\eqn\phissbeh{ e^\phi\simeq \left( {\pi\over 2 \ln r_0}\right)^2
{1\over |z|^2 \sin^2 ({\pi\over 2} {\ln |z|\over \ln r_0})} }  for
$|z|\ll1$. This is determined by rewriting the metric \tuba\ in the
coordinate $z=e^{(w-\pi)/l}$. From the integral of
$|\partial_z\phi|^2$ near $C$ and $C'$, as well as the boundary
term, the Liouville action behaves as \eqn\ror{\eqalign{ S_L &\sim
\int_{r_0} r dr (\partial_r \phi(r))^2 -4\phi(r_0)-8 \ln r_0  \cr
&\sim -4\ln r_0+{\cal O}(1).  }} Note that the length of the short
geodesic is \eqn\asitl{ 2\pi l = 2\pi r_0e^{\phi(r_0)/2} =
{\pi^2\over -\ln r_0} } Therefore we have $S_L \sim 2\pi/l + {\cal
O}(1)$ in the $l\to 0$ limit. This precisely cancels the singular
term from \asrft, and hence contribution the holomorphically
factorized partition function \szero\ remains finite in the $l\to 0$
limit, consistent with the expected factorization \zzf.

Let us write the full gravity partition function as $Z=Z_\gamma +
\check Z_{\gamma}$, where $Z_\gamma$ is the contribution from all
hyperbolic three-manifolds $M$ that fill in $\gamma$, and $\check
Z_{\gamma}$ is the contribution from the remaining gravitational
instantons, namely the ones such that $\gamma$ is not contractible
in $M$. By our conjectured relations \zfakedef, \fakedisc, as
$\gamma$ is pinched, $Z_\gamma$ factorizes on the Virasoro
descendants of the identity; if the dual CFT is extremal, this means
that $Z_\gamma$ already factorizes ``correctly" on states of
dimension $\Delta\leq k$, since all such states are Virasoro
descendants of the identity. To avoid spoiling this factorization,
one expects $\check Z_{\gamma}$ to contribute only to the
factorization on states of dimension $\Delta\geq k+1$. The above
discussion indicates that $\check Z_{\gamma}$ can only factorize on
states with $\Delta\geq k$, which is consistent with the dual CFT
having no nontrivial primaries up to dimension $k-1$. It is
intriguing whether the contribution to the factorization on
dimension $\Delta=k$ states in $\check Z_{\gamma}$ exactly cancel.
This would require cancellation say between certain handlebody
(subleading in the $Sp(2g,{\bf Z})$ Poincar\'e series of \YinGV) and
non-handlebody contributions in the pinching limit. This issue is
currently under investigation.

\newsec{Summary}

We have given a prescription for computing the classical contribution from
all hyperbolic instantons, handlebody or not, to the holomorphically factorized
partition function on a general Riemann surface. The $1/k$ quantum corrections to the contribution
from (non-handlebody) ``tight" manifolds remain to be understood. Once these are known, the
gravity partition function is in principle determined completely. In the end, we would like to
check the non-handlebody contributions against the dual ECFT, say by
examining the factorization on states and extracting correlation functions in the CFT.
It is also important to understand whether the gravitational instantons with multiple boundary components
can be consistently excluded.
These are left to future works.

\bigskip

\centerline{\bf Acknowledgement}
I would like to thank S. Giombi, D. Jafferis, B. Mazur, O. Saremi, A. Strominger, A. Tomasiello, and especially
D. Gaiotto and A. Maloney, for very useful discussions. I'm grateful to C. McMullen for answering
my questions on hyperbolic three-manifolds.
I am supported by a Junior Fellowship from the Harvard Society of Fellows.

\bigskip

\listrefs

\end

***********unused stuff *********

Let us consider an explicit example of a non-handlebody, obtained by
taking the ${\bf Z}_2$ quotient of $I\times \Sigma$, with hyperbolic
metric \eqn\hysaa{ ds^2 = dr^2 + \cosh^2r ds_\Sigma^2 } where
$ds_\Sigma^2$ is a hyperbolic metric on $\Sigma$. We shall consider
the case where $\Sigma$ has genus two, with period matrix $\Omega =
\pmatrix{\rho & \nu \cr \nu & \sigma }$. Restricting to the real
locus of its complex structure moduli space $\rho=-\bar\sigma$,
$\nu=i\nu_2$, $\nu_2\in{\bf R}$, $\Sigma$ admits an orientation
reversing fixed-point free involution. The ${\bf Z}_2$ action
$\iota$ on $I\times \Sigma$ is given by $r\to -r$ together with the
involution on $\Sigma$. According to section 3, we have \eqn\spsf{
2{\rm Re}S_0(\rho,\sigma=-\bar\rho,\nu=i\nu_2) =12\ln
{\zeta'_\Sigma(1)\over \rho_2^2-\nu_2^2} +const  } In the limit
$\nu_2\to 0$, $\Sigma$ pinches into two punctured tori connected by
a thin tube. In order to compute the Selberg zeta function in this
limit, we need to express the length $2\pi l$ of the short geodesic
around the tube in terms of the periods.

*****to be corrected*****

 We have \eqn\szetap{ \eqalign{ & e^{kS_0}
\sim e^{-4\pi k \rho_2} F(\rho,-\bar\rho,\nu_2). } } where $f$ and
$F$ are generic non-singular functions in the pinching limit. After
analytic continuation, we have \eqn\sans{ e^{kS_0}\sim e^{2\pi i
k(\rho+\sigma)} F(\rho,\sigma,\nu) } Once again, the contribution to
the partition function only factorizes on states of dimension
$\Delta\geq k$. We see that $e^{kS_0(M;\Sigma)}$ for $M$ a twisted
$I$-bundle only contributes to the factorization of the partition
function on states of dimension $\geq k$.

In the $\rho\to \infty$ limit, we have \eqn\aee{ \eqalign{ & w\sim
2e^{-\rho}+il \phi, \cr & ds^2 \sim d\rho^2 + l^2{e^{2\rho}\over
4}d\phi^2. } } To lead order approximation we can identify $\bar
z\sim e^{w/l}$ as the anti-holomorphic coordinate on the punctured
torus at the $\rho\to \infty$ end of the tube. In the $\rho\to
-\infty$ limit, we have \eqn\bee{ \eqalign{ & w\sim \pi-2e^{\rho}+il
\phi, \cr & ds^2 \sim d\rho^2 + l^2{e^{-2\rho}\over 4}d\phi^2. } }
We can identify $z'\sim e^{(w-\pi)/l}$ as the holomorphic coordinate
on the other punctured torus. Note that the two punctured tori are
oriented oppositely. Consider $\vartheta=de^{(w-\pi)/l}$ as a
holomorphic 1-form on the tube, extended as a harmonic form to all
of $\Sigma$. $\vartheta'\sim e^{-\pi/l}d\overline{z}$ on one torus
and $\vartheta'\sim d{z}'$ on the other torus. Similarly,
$\vartheta=de^{-w/l}$ is another holomorphic 1-form on $\Sigma$; it
approaches $d{z}$ on one torus and $e^{-\pi/l}d\overline{z}'$ on the
other. A normalized basis of holomorphic 1-forms,
$(\omega_1,\omega_2)$, is chosen as \eqn\omegaaa{ \omega_1 =
{\vartheta-e^{-\pi/l}\vartheta'\over 1-e^{-2\pi/l}},~~~~~~ \omega_2
= {\vartheta'-e^{-\pi/l}\vartheta\over 1-e^{-2\pi/l}}. } It follows
that ($\rho=-\bar\sigma$) \eqn\aart{ \nu_2 = 2e^{-\pi/l}\rho_2 +
{\cal O}(e^{-2\pi/l}). }

We can write the Selberg zeta function $\zeta_\Sigma(s)$ as
\eqn\contrsh{ \zeta_\Sigma(s)=\tilde f_\Sigma(s)\prod_{m=0}^\infty
(1-e^{-2\pi l(m+s)})^2, } where we singled out the contributions
from the short geodesics (counted with both orientations). $\tilde
f_\Sigma(s)$ denotes the contribution from all other closed
geodesics on $\Sigma$. $\tilde f_\Sigma(s)$, just like
$\zeta_\Sigma(s)$, has a simple zero at $s=1$. This can be
understood from the fact that the density of closed geodesics of
length $L$ grows as $\rho(L)\sim e^{-L}/L$ \Bogomolny. Therefore we
have \eqn\dgoa{ \zeta_\Sigma'(1) = \tilde
f_\Sigma'(1)\prod_{m=1}^\infty (1-e^{-2\pi ml})^2.  } Now ${\tilde
f_\Sigma}'(1)$ is finite in the $l\to 0$ limit, since it does not
involve the contribution from the short geodesic of length $2\pi l$.

Using \eqn\etafuf{ \prod_{m=1}^\infty (1-e^{-2\pi ml}) = e^{\pi
l\over 12}\eta(il) = {e^{\pi l\over 12}\over \sqrt{l}}\eta(i/l) \sim
e^{-{\pi\over 12l}}~~~~~(l\to 0) } we find \eqn\asrft{
\zeta_\Sigma'(1) \sim ({\nu_2\over \rho_2})^{1\over 6}
f(\rho,\bar\rho),~~~~\nu_2\to 0. } And hence the contribution to the
partition function scales like \eqn\eees{ e^{kS_0} \sim \nu_2^k
F(\rho,-\bar\rho),~~~~\nu_2\to 0, ~~\rho=-\bar\sigma. } After
analytic continuation to the Siegel upper half space, we have
\eqn\eesscont{ e^{kS_0} \sim \nu^k F(\rho,\sigma) + {\cal
O}(\nu^{k+1}) } In particular, it only contributes to the
factorization of the genus two partition function into torus
one-point functions of operators of dimension $\Delta\geq k$.